\input harvmac
\input epsf

\skip0=\baselineskip
\divide\skip0 by 2
\def\tmpsp{\the\skip0}

\def\skipthis#1{{}}

\Title{\vbox{\baselineskip12pt\hbox{hep-th/0002176}
\hbox{HUTP-00/A003}}}
{\vbox{\centerline{Domain Wall Junctions}
	\vskip2pt\centerline{in Supersymmetric Field Theories in $D=4$}}}
  
\centerline{Soonkeon Nam\foot{Permanent Address :
Dept. of Physics, Kyung Hee University; Seoul, 130-701, Korea, 
{\tt nam@string.kyunghee.ac.kr}}
and Kasper Olsen} 

\bigskip\centerline{\it Department of Physics}
\centerline{\it Harvard University}
\centerline{\it Cambridge, MA 02138}
\centerline{\tt nam@pauli.harvard.edu, kolsen@feynman.harvard.edu}

\vskip .3in

We study the possible BPS domain wall junction configurations for general
polynomial superpotentials of ${\cal N}=1$ supersymmetric Wess-Zumino 
models in $D=4$.
We scan the parameter space of the
superpotential and find different possible BPS states for different values
of the deformation parameters and present our results graphically.
We comment on the domain walls in F/M/IIA theories obtained 
from the Calabi-Yau fourfolds with isolated singularities and a background flux.

\smallskip
\Date{02/00}
\lref\rs{L. Randall and R. Sundrum, Phys. Rev. Lett.
{\bf 83} (1999) 3370; hep-th/9906064.}
\lref\cs{C. Cs\'aki and Y. Shirman, 
{\it Brane Junctions in the Randall-Sundrum Scenario}, Phys. Rev. {\bf D61} 
(2000) 024008, hep-th/9908186.}
\lref\nelson{A.E. Nelson, hep-th/9909001.}
\lref\network{Soonkeon Nam, {\it Modeling a Network of Brane Worlds,}
hep-th/9911104.}
\lref\oins{H. Oda, K. Ito, M. Naganuma, and N. Sakai, 
{\it An Exact Solution of BPS Domain Wall Junction}, hep-th/9910095.}
\lref\cv{S. Cecotti and C. Vafa, {\it On Classification of ${\cal N}=2$
Supersymmetric Theories}, Commun. Math. Phys. {\bf 158} (1993) 569.}
\lref\manyfold{N. Arkani-Hamed, S. Dimopoulos, G. Dvali, and N. Kaloper,
hep-th/9911386.}
\lref\hstt{H. Hatanaka, M. Sakamoto, M. Tachibana, and K. Takenaga,
hep-th/9909076.}
\lref\ab{E.R.C. Abraham and P.K. Townsend, {\it Intersecting Extended
Objects in Supersymmetric Field Theories}, Nucl. Phys. {\bf B351} (1991) 313.}
\lref\fmvw{P. Fendley, S.D. Mathur, C. Vafa, and N.P. Warner, 
{\it Integrable Deformations and Scattering Matrices for the ${\cal N}=2$ 
Supersymmetric Discrete Series}, Phys. Lett. {\bf 243B}
(1990) 257.}
\lref\bpsjunc{G.W. Gibbons and P.K. Townsend, {\it A Bogomol'nyi Equation
for Intersecting Domain Walls}, 
Phys. Rev. Lett. {\bf 83} (1999) 1727.}
\lref\gght{J.P. Gauntlett, G.W. Gibbons, C.M. Hull, and P.K. Townsend, {\it
BPS states of $D=4$ ${\cal N}=1$ supersymmetry}, hep-th/0001024.}
\lref\cht{S.M. Carroll, S. Hellerman, and M. Trodden, {\it Domain Wall
Junctions are 1/4--BPS states}, to appear in Phys. Rev. D, hep-th/9905217.}
\lref\chttwo{S.M. Carroll, S. Hellerman, and M. Trodden, 
{\it BPS Domain Wall Junctions in Infinitely Large Extra Dimensions }, 
hep-th/9911083.}
\lref\phrem{P. K. Townsend, {\it PhreMology: Calibrating M-branes},
hep-th/9911154}
\lref\gvw{S. Gukov, C. Vafa, and E. Witten, 
{\it CFT's From Calabi--Yau Four--Folds}, 
hep-th/9906070.}
\lref\dgeometry{See for example M.R. Douglas, {\it Topics in D-Geometry}, 
hep-th/9910170, and references therein.}
\lref\sv{A.D. Shapere and C. Vafa, {\it BPS Structure of Argyres--Douglas
Superconformal Theories}, hep-th/9910182.}
\lref\bg{K. Behrndt and S. Gukov, {\it Domain Walls and Superpotentials
from M theory on Calabi--Yau three--folds}, hep-th/0001082.}
\lref\mqcd{E. Witten, {\it Branes and the Dynamics of QCD}, hep-th/}
\lref\dk{G. Dvali and Z. Kakushadze, 
{\it Large N Domain Walls as D-branes for ${\cal N}=1$ QCD String}, 
Nucl. Phys. {\bf B537} (1999) 297, hep-th/9807140.}
\lref\gukov{S. Gukov, {\it Solitons, Superpotentials and Calibrations}, hep-th/9911011.}
\lref\liquidcrystal{See for example M. Kl\'eman, {\it Points, Lines and Walls}, 
John Wiley and Sons, New York, 1983.}
\lref\ttbar{S. Cecotti and C. Vafa, {\it Topological--anti--topological
Fusion}, Nucl. Phys. {\bf B367} (1991) 359.}
\lref\binosi{D. Binosi and T. ter Veldhuis, Domain wall junctions in a 
generalized Wess-Zumino model, hep-th/9912081.}
\lref\bazeia{D. Bazeia and F.A. Brito, Bags, junctions and networks of BPS and non-BPS defects, hep-th/9912015.}
\lref\gabadadze{G. Gabadadze and M. Shifman, {\it D-Walls and Junctions in 
Supersymmetric Gluodynamics in the Large $N$ Limit Suggest the Existence of 
Heavy Hadrons}, hep=th/9910050.}
\lref\hananywitten{A. Hanany and E. Witten, {\it Type IIB superstrings, 
BPS Monopoles, and Three Dimensional Gauge Dynamics,} Nucl. Phys. {\bf B492}
 (1997) 152, hep-th/9611230.}
\lref\vafatalk{C. Vafa, talk at ``String theory at the Millenium''
Conference, Caltech Jan. 2000, \ \ \   
{\tt http://quark.theory.caltech.edu/people/rahmfeld/Vafa/fs1.html};
K. Hori, A. Iqbal,  and  C. Vafa, to appear.}
\lref\cosmology{See for example A. Vilenkin and E.P.S. Shellard , 
{\it Cosmic Strings and Other Topological Defects},
Cambridge University Press, Cambridge, 1994.}
\lref\argyresdouglas{P.C. Argyres and M.R. Douglas,
{\it New Phenomena in $SU(3)$ Supersymmetric Gauge Theory,}  
Nucl. Phys. {\bf B448} (1995) 93, hep-th/9505062.}
\lref\apsw{P.C. Argyres, M.R. Plesser, N. Seiberg, and E. Witten,
{\it New ${\cal N}=2$ Superconformal Field Theories in Four Dimensions}
Nucl. Phys. {\bf B461} (1996) 71, hep-th/9511154.}
\lref\kazamasuzuki{Y. Kazama and H. Suzuki,
{\it New ${\cal N}=2$ 
Superconformal Field Theories and Superstring Compactificaton,}
Nucl. Phys. {\bf B321} (1989) 232.} 
\lref\ehiy{T. Eguchi, K. Hori, K. Ito, and S.-K. Yang,
{\it Study of ${\cal N}=2$ Superconformal Field Theories in Four Dimensions},
Nucl. Phys. {\bf B471} (1996) 430, hep-th/9603002.}
\lref\polstrom{J. Polchinski and  A. Strominger, {\it New Vacua for Type II
String Theory,} Phys. Lett. {\bf B388} (1996) 736, hep-th/9510227.}
\lref\beckerbecker{K. Becker and M. Becker, 
{\it  M Theory on Eight Manifolds,}
Nucl. Phys. {\bf B477} (1996) 155, hep-th/9605053.}
\lref\sethivafawitten{S. Sethi, C. Vafa, and  E. Witten, 
{\it Constraints on Low Dimensional String Compactifications,}
Nucl. Phys. {\bf B480} (1996) 213, hep-th/9606122.}
\lref\dasgupta{K. Dasgupta, G. Rajesh, and S. Sethi, 
{\it M Theory, Orientifolds and $G$-Flux},
JHEP {\bf 9908} (1999)
023, hep-th/9908088.}
\lref\MQCD{E. Witten, {\it Branes And The Dynamics Of QCD} 
Nucl. Phys. {\bf B507} (1997) 658, hep-th/9706109.}
\lref\OoguriVafa{H. Ooguri and C. Vafa, {\it Two Dimensional Black Hole and
Singularities of CY Manifolds,} Nucl. Phys. {\bf B463} (1996) 55, hep-th/9511164.}
\lref\GHM{R. Gregory, J. Harvey, and G. Moore, {\it Unwinding Strings and
T-Duality of Kaluza-Klein and H Monoples}, Adv. Theo. Math. Phys., {\bf 1} (1997) 283, hep-th/9708086.}
\lref\KLMV{A. Klemm, W. Lerch, P. Mayr, C. Vafa, and N. Warner,
{\it Self-Dual Strings and ${\cal N}=2$ Supersymmetric
Field Theory}, Nucl. Phys. {\bf B477} (1996) 746, hep-th/9604034.}
\lref\junctionpapers{ P.M. Saffin,{\it Tiling with almost BPS Junctions},
Phys. Rev. Lett. {\bf 83} (1999) 4249, hep-th/9907066;
D. Bazeia and F.A. Brito, {\it Tiling the Plane without Supersymmetry}, 
hep-th/9908090;
A. Gorsky and M. Shifman, {\it More on the  Tensorial central charges in 
${\cal N}=1$ Supersymmetric Gauge Theories (BPS Wall Junctions and Strings)}, 
hep-th/9909015;
D. Bazeia and F.A. Brito, {\it Bags, Junctions, and Networks of BPS and nonBPS
 Defects}, hep-th/9912015;
D. Binosi and T. ter Veldhuis, {\it  Domain Wall Junctions in a generalized
Wess-Zumino Model}, hep-th/9912081. }
\lref\kachruvafa{S. Kachru and C. Vafa, 
{\it Exact results for ${\cal N}=2$ Compactification of Heterotic
Strings,}, 
Nucl. Phys. {\bf 450} (1995) 69, hep-th/9505105;
A. Klemm, W. Lerche, P. Mayr, and C. Vafa, {\it
Nonperturbative results on the point particle limit of ${\cal N}=2$
Heterotic String Compactification}, 
Nucl. Phys. {\bf B459} (1996) 537, hep-th/9508155.}
\lref\witten{E. Witten, {\it Non-Perturbative Superpotentials 
In String Theory}, Nucl.Phys. {\bf B474} (1996) 343, 
hep-th/9604030.}

\newsec{Introduction}

Domain walls arise in scalar field theories as solutions connecting two
isolated vacua which are degenerate.
Physical examples can range from a system of liquid crystals \liquidcrystal\ 
to defects in cosmological models \cosmology.
A simple way to obtain a theory with degenerate vacua  is to consider 
a supersymmetric field theory.
In this case supersymmetry guarantees 
the positivity of the scalar potential $V(\Phi)$, which can 
be written in terms of superpotential $W(\Phi)$, i.e. $V(\Phi)\sim 
\left|{\partial W(\Phi)\over \partial \Phi}\right|^2$.
The location of the minima of the potential are at the critical points
$\Phi=\Phi_k$ 
of the superpotential, such that $W'(\Phi_k)=0$. 

Starting from the simplest model of a single scalar field theory with a
potential in $1+1$ dimensions, which allows a single type of 
domain wall between each of the critical points, 
things get more complicated when we consider theories with multiple 
scalar fields and multiple critical points.
In cases where there are more than two degenerate vacua, 
one might consider any pair of 
vacua and try to connect them with a domain wall 
(or soliton in $1+1$ dimension).
However, this simple--minded construction cannot always be realized since
there might not always be a BPS solution connecting two given vacua.
This can be exemplified by the Wess-Zumino(W-Z) model with 
the following quintic superpotential:
$W(\Phi) = \Phi^5/5 - \Phi^2/2$, which
has  four critical points, one at $\Phi=0$ and three
others at vertices of an equilateral triangle.
In this theory the domain wall which interpolates between 
$\Phi=0$ and any one of the
corners exists, but direct connection of two of the vertices does 
not exist \fmvw.
Therefore such a superpotential only allows for three BPS states and not
six as one might have expected.
(For this particular example, one can actually see from surface plot of 
the potential $V(\Phi)$ that there is no BPS path between the vertices of the 
triangle).

In $1+1$ dimensions these interpolating BPS solutions are just kinks or solitons. 
Integrability conditions for different soliton solutions in $1+1$
dimensions, interpolating 
different pairs of critical points were studied in Ref.\fmvw, where 
a soliton which saturates the Bogomol'nyi bound 
can best be described as a straight line connecting the critical points in the
superpotential space, i.e. the $W-$plane.
In fact a very extensive classification program of integrable models
was carried out in $1+1$  dimensional theories with ${\cal N} =2$ 
supersymmetry in Ref.\cv.
Some of the results there can be used in higher--dimensional theories with
domain walls because domain walls essentially have one space dimensional 
dependence, which is along the direction separating two domains.
One new feature that appears when we have more than one 
spatial dimension is that 
we can now have intersections or junctions of domain walls \ab.
We can ask a similar question for the existence of a BPS state between
critical points each time we encounter a superpotential, and perform an
analysis as was done extensively in Ref.\ttbar.
However, it would be desirable to have a more global view in the parameter
space (i.e. the space of deformations of the superpotential) 
so that we can easily follow the behavior of certain BPS states which are 
created or destroyed as we move around in this parameter space.

In this paper, we will consider domain walls and their junctions in 
${\cal N}=1$ supersymmetric field theories in four dimensions and we
analyze under which circumstances certain classes of junctions can appear
or not.   
For an appropriate choice of superpotential, such domain walls have been 
shown to arise in the W-Z model and also in $SU(N)$ SUSY QCD for which the W-Z
model is an effective low--energy theory. Furthermore, it has been shown
that the W-Z model (at least for a ${\bf Z}_3$ symmetric configuration of
three critical points) admits solutions preserving only 1/4 of
supersymmetry ~\bpsjunc\cht\phrem, which were interpreted as 
junctions of three domain walls. More general BPS and non-BPS junctions of
the W-Z model with a ${\bf Z}_k$ symmetric configuration of critical points
where discussed in \junctionpapers.
Recently nonperturbative junctions of domain wall solutions were also extensively
studied in SUSY QCD ~\dk, and in the brane world scenarios
\cs\chttwo\network, where 
gravitating domain wall junctions were considered.

Another important motivation to study this subject comes from the recent
discussions of the vacuum and soliton structure of
supersymmetric theories in the context of string
theory compactifications \polstrom\gvw. 
Consider compactification
of Type II, $M$-, or $F$-theory on some singular noncompact 
Calabi--Yau $n$ manifold with some background flux of Ramond-Ramond field,
say $G$. (For $F$-theory, we need elliptically fibered Calabi-Yau manifold,
and in addition we need both NS and RR fluxes.)
Nonvanishing R-R flux is needed to cancel the tadpole anomaly \sethivafawitten,
while taking a singular limit of a Calabi-Yau manifold leads to a decoupling of 
gravity in the effective field theory in the lower dimension \kachruvafa .
Domain walls are identified with
D-branes (or M-branes for M-theory) 
wrapped on supersymmetric cycles and in crossing such a brane
the flux (of the appropriate field) jumps, so the different values of the
flux correspond to different vacua. For supersymmetric vacua certain
conditions has to be imposed on $G$ \beckerbecker. 
These constraints can be realized 
by interpreting $G$ as giving rise to an effective superpotential 
of the lower--dimensional theory
which is of the form 
\eqn\gpot{W=\int A\wedge G,}
where $A$ is either the holomorphic $n$-form $\Omega$ 
or some appropriate power of the K\"ahler potential ${\cal K}$.
\foot{Note that this is related to the theory of 
{\it calibrations}: 
$A$ is the calibration and
for $A=\Omega$ these potentials are related to Lagrangian
submanifolds and give rise to chiral superfields, while if 
$A={\cal K}^p$ they
are related to holomorphic curves and lead to ``twisted'' chiral
superfields. \dgeometry }     
For compactification of Type II, $M$-theory or $F$-theory on 
{\it singular} Calabi--Yau manifolds this analysis leads in certain cases 
to an identification of the corresponding low--dimensional
theories as specific non--trivial conformal field theories, 
depending on the singularity in question. As an example, 
it was shown \gvw\ that
Type IIA compactified on a Calabi--Yau four-fold with $A_n$ singularity
gives an ${\cal N}=2$ Kazama-Suzuki model \kazamasuzuki\ in two 
dimensions.

In this paper, we will concentrate on W-Z models in four dimensions (with
four supercharges), though much of the analysis can be applied in three and
two dimensions as well.
We analyze the appearance of BPS domain walls and junctions for massive
deformations away from the conformal point.
In section 2, we review the possibility of central charges of the 
${\cal N}=1$
superalgebra in four dimensions and their interpretation in terms of domain
wall and junction charges and also 
the BPS condition for the domain walls and their
junctions. In section 3 we review the derivation of W-Z models 
in $D=2,3$ from type IIA or M-theory 
and discuss some relations between the geometry of
the Calabi--Yau manifold and the solutions of the BPS equation in lower 
dimensions. We also comment about generating superpotentials in $F$-theory.
In section 4, we collect the rules for the counting of BPS states, which
are used in section 5 in studying massive deformations of the W-Z model with
a general quintic superpotential. Finally, section 6 contains our 
discussions.
 
\newsec{Supersymmetry Algebra and the BPS Condition}

We start by recalling the structure of the 
${\cal N}=1$ supersymmetry algebra in $3+1$ dimensions and how the possibility 
of domain walls  and junctions of domains walls can be analyzed 
directly from this algebra.
(For further discussions of the ${\cal N}=1$ algebra in $D=4$, 
see \bpsjunc\cht\oins\gght).

\lref\agit{J.A. de Azc\'arraga, J.P. Gauntlett, J.M. Izquierdo, and
P.K. Townsend,
{\it Topological Extensions of the Supersymmetry Algebra for Extended Objects.}
Phys. Rev. Lett. {\bf 63} (1989) 2443.} 

The ${\cal N}=1$ supersymmetry algebra in $D=4$ allows
central charges which correspond to tensions of BPS domain walls and junctions
of them \agit\oins:
\eqn\susyalg{\eqalign{
& \{  Q_\alpha, Q_\beta \} = 2i (\sigma^k \overline{\sigma}^0)_
\alpha^{\ \gamma}\epsilon_{\gamma\beta} Z_{k},\cr
&  \{  Q_\alpha, \overline{Q}_{\dot{\alpha}}\} = 
2 (\sigma^{\mu}_{\alpha\dot{\alpha}}P_\mu + \sigma^k_{\alpha\dot{\alpha}}
Y_k),}}
where $k=1,2,3$ and $\mu=0,\ldots,3$.
The $Z_k$ (which are complex charges) have an interpretation as 
domain wall charges and $Y_k$ (which are real charges) as the junction
energy, which can be either positive or negative \oins.

The relations between the superpotential and the central charges are given by
\eqn\centralchargeZ{
Z_k = 2\int d^3x \partial_k W^*(\phi^*)}
\eqn\centralchargeY{
Y_k = i\epsilon^{knm}{\int d^3x  K_{i\bar{j}}\partial_n(\phi^{*j}
\partial_m \phi^i)}.}
where $\phi$ is the scalar component of the chiral superfield, and 
the K\"ahler metric is derived from the K\"ahler potential $K$ via 
$K_{i\bar{j}}=\partial^2K/\partial\phi^i\partial\phi^{*j}$.
The central charges $Z_k$ depend only on the difference between 
the values of the superpotential at spatial infinity. 
If we have a single domain wall 
-- which is only nontrivial in one dimension --
then $Y_k$ vanishes for all $k$ and $Z_j$ is nonvanishing (for some $j$) 
since the $Z_j$ central charge depends on the spatial derivative in the
$x_j$ direction. For junctions of domain walls to be possible, one first of
all need more than a single (real) scalar field, since else $Y_k$ will vanish
identically. Furthermore, as one can clearly see, $Y_k$ is 
nonvanishing only when the field configuration
at infinity is nontrivial in two dimensions.
If we have two spatial dependences as for a domain wall junction solution, 
then we will generically have two of the $Z_k$'s nonvanishing.
When the K\"ahler metric is trivial, $Y_k$ is just a surface integral
at the infinity. 

When we start from a ${\cal N}=1$ theory in $D=4$ we originally have four
supercharges.
Domain wall configurations with nonzero $Z_k$'s and vanishing $Y_k$ 
has two conserved supercharges, thus are 1/2 BPS states, 
whereas when there is nonzero $Y_k$ there is only one combination of the four
supercharges which can survive. This leads to a 1/4 BPS state. 

The BPS equation for a single static domain wall of the W-Z model 
(dimensionally reduced to two dimensions) is given by:
\eqn\BPS{\partial_x \phi = e^{i\alpha}\overline{W'}}
where the prime denotes derivative with respect to $\phi=\phi(x)$, 
and $x$ is a coordinate and $\alpha$ is an arbitrary phase. 
A domain wall solution of mass $M$ saturates
the bound $M\geq |T|$, where $T$ is the topological charge associated with
the wall and has $\alpha=\arg T$.  
The BPS equation for a domain wall junction can be derived in higher
space dimension as in \bpsjunc\ 
and is completely analogous to Eq.\BPS.
In particular, if we suppress spatial dependences other than two of them,
say $x$ and $y$, then the BPS equation becomes 
\eqn\BPSj{{(\partial_x-i\partial_y) \phi = e^{i\alpha}\overline{W'}}.}
The BPS solution saturates the bound $M\geq |T|+Q$, where $Q$
is the junctions charge.
When there is only on spatial dependence, e.g. $\partial_y\phi(z)=0$, 
this reduces to Eq. \BPS.
Note that this junction is an object in a three-dimensional theory and not
a two-dimensional theory as the one discussed in \bpsjunc.

\newsec{Wess-Zumino Models from Calabi-Yau Compactifications}
\lref\kutasov{See for example Sec. V.D.2 of A. Giveon and D. Kutasov,
{\it Brane Dynamics and Gauge Theory}, Rev. Mod. Phys.  {\bf 71} (1999) 983,
hep-th/9802067.}

The W-Z model we will consider will be a field theory in $D=4$ 
dimensions with ${\cal N}=1$ supersymmetry.
It has a superpotential $W(\Phi)$ which is of the form
\eqn\superpotential{W'(\Phi) = C \prod^m_{i=1}( \Phi-\lambda_i),}
where $C$ is a constant and $\lambda_i$, $i=1,\ldots, m$ 
are the locations of the critical points in the field
space. The ${\bf Z}_m$ symmetric case, for
$\lambda_i = |\lambda_0|e^{2\pi i /m}, (i = 1, \cdots, m),$
were considered in connection to the ${\cal N}=1$ 
supersymmetric $SU(N)$ YM theory in the large $N$ limit\dk. 
Although it is believed that W-Z models for $m>3$ will flow to trivial IR
theories for $D\geq 3$, it can become relevant as a perturbations to
certain fixed points. Furthermore one can have non-trivial brane
configurations realizing these higher order potentials \kutasov.
We will comment on this in the last section.

As discussed in \gvw\ Wess-Zumino models can arise in Calabi--Yau
compactifications of M/Type IIA theories.  
The locations of the isolated singularities correspond to the locations of 
the critical points. In the case of an 
$A_k$ singularity, the local
geometry of the Calabi--Yau $n$--fold is described by an equation of the
form 
\eqn\sing{-P_m(z_1)+z_2^2+\ldots z_{n+1}^2=0} 
with $P_m(z_1)$ a generic polynomial of degree $m=k+1$ in $z_1$:
\eqn\polynomial{P_m(z_1)= \prod_i (z_1-a_i).}
In the above $a_i$ are the locations of the singularities.
When we fix $z_1$ we can regard $|P_m(z_1)|$ as the radius of the $n-1$ sphere
which is the nontrivial cycle of the manifold.
Since the mass of the brane wrapping around the singularity will be proportional
to the volume of the sphere, and this mass will give also the tension of the 
domain wall in the  lower effective field theory, we get 
the following fact that the
superpotential $W$ is related to $P_m$ through
\eqn\cypot{dW = P^{{n-2}\over2}_m dz_1,}
where the right hand side is basically the volume of the sphere.
So, for Calabi-Yau fourfold compactifications we recover the W-Z superpotentials.

Instead of going directly to the case of a quintic superpotential, we would
like to discuss some general features of the solutions of the BPS equation
corresponding to compactification on a general Calabi--Yau $n$-fold. 
BPS states are in this case identified with wrapped $n$-branes in a
Calabi--Yau $n$-fold near an isolated singularity \gvw.
The kind of singularities we will be looking at are the $A_k$
singularities, which are described by the Eq.\sing. Any such Calabi--Yau manifold
has a unique $n$-form $\Omega$ which determines the volume of a cycle
$C$. The condition for a cycle to be minimal is that its volume saturates
the inequality
\eqn\min{V=\int_C|\Omega|\geq\left|\int_C\Omega\right|.}
Now, the problem of minimizing this volume can be mapped to a problem in
the complex $z_1$-plane as follows. One considers the
$n$--cycle to be an $n$--sphere $S^{n}$, which is locally of the form 
$S^{n-1}\times S^1$, i.e. as an $S^{n-1}$--sphere fibered over a real curve in the
$z_1$--plane. The local volume of this $S^{n-1}$--sphere is determined by
$z_2^2+\ldots z_{n+1}^2=P_m(z_1)$ and so vanishes at the roots of
$P(z_1)$, which are identified with the critical points of the superpotential
$W(z_1)$. 
With this choice of local coordinates on the singular
Calabi--Yau, the expression for the holomorphic $n$-form is
\eqn\holm{\eqalign{\Omega = &{dz_1\cdots dz_n \over z_{n+1}} \cr  
= & {idz_1\cdots dz_n \over \sqrt{z_1^2+\cdots +z_n^2-P_m(z_1)}}}}  
The condition for a cycle to be supersymmetric is that 
the image of the path is a straight line in the flat $W$--plane, 
where $W$ is defined through the relation in Eq. \cypot. 
This comes from minimizing the l.h.s. of \min\ with 
the expression \holm\ for $\Omega$. The BPS 
condition is then:
\eqn\stline{W(z(t))=\int^{z(t)}_{z_0}P^{n-2 \over 2}dz=\alpha t,}
where $t$ parametrizes the curve connecting the two critical points.
To obtain the BPS states one should therefore solve the
first--order differential equation:
\eqn\intcurve{P^{n-2 \over 2}{dz\over dt}=\alpha}
with the boundary conditions that $z(t)$ should begin and end at the roots
of $P(z)$ (or rather of $dW(z)$). Near a root, which we take to be at $z=0$,
one is solving an equation of the form
\eqn\root{{dz\over dt}={\alpha\over z^{n-2 \over 2}},}   
for which the solution is 
\eqn\rootsol{z=\left({n\over 2}\alpha t\right)^{2/n}.}
In the case of a Calabi--Yau
four--fold we see that there are four solutions for any $\alpha$ and that
the corresponding curves intersect at an angle of 90$^\circ$.

Now we will discuss how to construct domain walls in such Calabi--Yau
compactifications 
and we will follow the discussion in \gvw. 
For more details, see also \dasgupta\bg.
Consider compactification of $M$-theory on some Calabi--Yau 
four--manifold $Y$
with some background flux of the three--form potential $C$ which couples to
the membranes
(see Table 1, in which we summarize the construction of vacua and domain
walls in compactification of M/IIA/$F$-theory with background fluxes as in \gvw).
These $C$-field are classified by a class $\xi\in H^4(Y;{\bf Z})$, which in
turn defines a lattice $\Gamma^*=H^4(Y;{\bf Z})$. 
This set of data specifies a choice of vacuum. 
Now, to make a domain wall 
in ${\bf R}^3$, one considers a fivebrane with worldvolume ${\bf R}^2\times
S$, where $S$ is a four-cycle. Such four--cycles are classified by 
$H_4(Y;{\bf Z})$, which defines a lattice $\Gamma$ dual to $\Gamma^*$. 
So when crossing such a domain wall $\xi$ changes -- and the possible values of
$\xi$ are classified by $\Gamma^*/\Gamma$. 
The effective superpotential is obtained as follows. 
When crossing a domain wall, the change in the superpotential is
\eqn\deltasup{\Delta W={1\over 2\pi}\int_X\Omega\wedge\Delta G,}
where $G=dC$.
Here $X={\bf R}^3\times Y$. This superpotential will then account for the 
restrictions on $G$ (which are implied by having vacua with supersymmetry) 
mentioned in the introduction.

Compactifying Type IIA on $Y$, we have $X={\bf R}^2\times Y$. And to
specify a vacuum we should also specify the topological class of the
$G$--field, which is now the RR four--form field and takes values in
$H^4(Y;{\bf Z})$. To make a domain wall,
one now has four--branes with worldvolume ${\bf R}\times S$. Again the possible
four--cycles $S$ are classified by $H_4(Y;{\bf Z})$ and therefore $\xi$ 
takes values in $\Gamma^*/\Gamma$. 
Compactification of Type IIA on Calabi--Yau four--fold $Y$, 
with $A_k$ singularity, will then give
an effective two-dimensional theory with superpotential determined by
\cypot\ for $n=4$. This is precisely the superpotential discussed in the following
sections, and here we can of course have domain walls between different
vacua. But we will not have junctions. 

\lref\ftheory{C. Vafa, {\it Evidence for $F$-theory}, 
Nucl. Phys. {\bf B469} (1996) 403, hep-th/9602022.}
\lref\taylorvafa{T.R. Taylor, C. Vafa,
{\it RR Flux on Calabi-Yau and Partial Supersymmetry Breaking,} 
to appear  Phys. Lett. B, hep-th/9912152.}
The story for $F$-theory \ftheory\
compactifications is slightly different.
First of all, 
for $F$-theory compactification on ${\bf R}^4\times Y$ we need $Y$ to be
an elliptically fibered four--manifold. The flux $\Phi$ discussed in \gvw\ now
has contributions from space--filling threebranes and not membranes as in
the compactification of $M$-theory. 
The analog of the $G$-field now becomes both 
$NS$ and $RR$ three--form fields, 
$H^{NS}$ and $H^R$, from the Type IIB theory. 
$F$-theory on ${\bf R}^4\times Y$ can be described as Type IIB with certain
$(p,q)$-sevenbranes on a locus $L\subset B$, where 
$B$ is the base of the elliptic fibration. 
However, in this situation,
one can find a simpler description:
This $F$-theory compactification with singularity can be
reinterpreted as Type IIB with a D7-brane with worldvolume ${\bf R}^4\times L$,
where $L$ is a complex (singular) surface inside ${\bf C}^3$ (see Table 1).
This specifies a choice of vacuum. One has a $U(1)$-gauge field on the
D7-brane and so this vacuum is characterized by the first Chern class, or
an element $\xi$ of the lattice $\Gamma^*=H^2(L;{\bf Z})$. 
\medskip
{\vbox{\ninepoint{
$$
\vbox{\offinterlineskip\tabskip=0pt
\halign{\strut
\vrule#
&
&\hfil ~$#$
&\hfil ~$#$
&\hfil ~$#$
&\hfil ~$#$ 
&\hfil ~$#$
&\hfil ~$#$ 
&\hfil ~$#$
&\hfil ~$#$
&\vrule#\cr
\noalign{\hrule}
&
&
&M{\rm -theory}
&  
&{\rm IIA-theory}
&  
&F{\rm -theory}
&  
&\cr
\noalign{\hrule}
&{\rm Vacuum}:
&
&{\bf R}^3\times Y
&
&{\bf R}^2\times Y
&
&{\bf R}^4\times Y
&
&\cr
&
&
&G=dC
&
&G=dC
&
&D7={\bf R}^4\times L
&
&\cr
&
&
&G\in H^4(Y;{\bf Z})=\Gamma^*
&
&G\in H^4(Y;{\bf Z})=\Gamma^*
&
&F\in H^2(L;{\bf Z})=\Gamma^*
&
&\cr
&
&
&
&
&
&
&
&
&\cr
&{\rm Domain\ wall}:
&
&M5={\bf R}^2\times S
&
&D4={\bf R}\times S
&
&D5={\bf R}^3\times V
&
&\cr
&
&
&[S]\in H_4(Y;{\bf Z})=\Gamma
&
&[S]\in H_4(Y;{\bf Z})=\Gamma
&
&[\partial V]\in H_2(L;{\bf Z})=\Gamma
&
&\cr
&
&
&
&
&
&
&
&
&\cr
&
&
&
&
&
&
&
&
&\cr
\noalign{\hrule}}\hrule}$$
\vskip -3 pt

\centerline{{\bf Table 1:}M/IIA/F--theory on Calabi--Yau four--fold $Y$.}
\vskip7pt}}} 
\bigskip

\noindent How do we construct domain walls?
Take a D5-brane, which can end
on the D7-brane, with worldvolume ${\bf R}^3\times V$, where $V$ is a
three--manifold whose boundary should be in $L$ (since the D5-brane ends on
the D7-brane). This boundary defines a topological class 
$[\partial V]\in H_2(L;{\bf Z})$, i.e. in the lattice 
$\Gamma=H^2(L;{\bf Z})$ dual to $\Gamma^*$. Crossing the domain wall, the
Chern class changes by the amount $[\partial V]$. Again $\xi$ takes values 
in $\Gamma^*/\Gamma$.
We also need to specify the local geometry of $Y$.
For elliptic four-fold singularity one has the description
\eqn\elliptic{y^2=x^3+3ax^2+H(z_1,z_2,z_3),}
where $H$ is a polynomial in $z_1,z_2,z_3$.
The equation for $L$ then becomes simply $H=0$ and to describe an 
$A_k$-singularity one
should then choose:
\eqn\Ak{H=z_1^{k+1}+z_2^2+z_3^2.} 
It would be desirable to have an explicit computation of the superpotential 
in $F$-theory generated by the inclusion of $H$-flux and with   
$A-D-E$--type singularities. For that one could start with Type IIB on
Calabi--Yau three--fold as in \taylorvafa, where 
$W=\int\Omega\wedge(\tau H^{NS}+H^R)$, and then lift this construction to 
$F$-theory. 
Note, however, that not all $Y$ will generate a 
nontrivial superpotential \witten.

\newsec{Rules for the Construction}

Now we will discuss the rules for finding the number of BPS states for
different values of the perturbation parameters, which translates to
varying the positions of the critical points.  

1) {\it What are we constructing?}

 From the BPS equation one can easily show that the BPS solution trajectories
are straight lines between critical points in the $W$--plane \cv. 
However the inverse image of a certain straight line -- connecting, say
$W(i)$ and $W(j)$ -- might not lift back to the field space as a curve 
connection the vacua and thus does not correspond to a BPS solution. 
To count the number of actual solutions connecting $i$ and $j$, one starts
with the ``wavefront''(or sphere) of all possible solutions emanating 
from $i$ with fixed values of $W$, denoted by $\Delta_i$, and the same 
for the critical point
$j$. 
\lref\susyindex{S. Cecotti, P. Fendley, K. Intriligator, C. Vafa, 
{\it  A New Supersymmetric Index,}
Nucl. Phys. {\bf B386} (1992) 405-452, hep-th/9204102.} 
The number of solutions is then exactly the number of points at which
$\Delta_i$ and $\Delta_j$ intersect \cv (note that the intersection number
depends on a choice of orientation and what we really are
computing is a weighted sum\susyindex ). This defines the intersection number 
$\mu_{ij}=\Delta_i\circ\Delta_j$ as a quantity which is invariant under
small perturbations of the superpotential since it is integer. 
However, as we vary the
superpotential the critical points will move around in the $W$--plane and
when a third root $k$ crosses the straight line connecting $i$ and $j$ the
number of BPS solutions connecting $i$ and $j$ can obviously
change. Precisely how this number changes can be derived using the
Picard-Lefschetz theorem and is given by \cv
 \eqn\pl{\mu'_{ij}=\mu_{ij}\pm \mu_{ik}\cdot \mu_{kj}}
Here the $\pm$ sign depends on the ordering of $ikj$ in the 
triangle defined by
the three roots before $k$ was crossing the line between $i$ and $j$. 
(Physically, this change in the intersection number, as one root crosses the
line between two other roots, can be understood in terms of the
Hanany-Witten effect \hananywitten\vafatalk.) 
In principle on can determine these intersection numbers by
solving the so--called tt$^*$ equation \ttbar\cv\ for a fixed choice of
superpotential. But in our case we vary the parameters in the
superpotential and it is more straightforward to look at conditions on
masses of BPS solutions (and phases of the topological charges) to
determine which kind of junctions exist or not. 
So we are in a certain sense trying to give a unified
description of the cases considered in Ref. \ttbar.

So, it would be nice to have the form of $\mu_{ij}$'s as functions of the 
parameters of the theory. Since it takes integer values, it is
stable under small perturbations and changes only by an integer, 
and the best way to represent it would be to find the 
boundaries in the parameter space
where the jump in the values happens. (This is called a {\it separatrix curve}.)
Then we can specify the values of 
$\mu_{ij}$'s inside each domain separated by the boundaries in the
parameter space graphically. In the case of W-Z models 
we have $|\mu_{ij}|=0$ or 1.
Crossing a boundary induces a change in the number of BPS state of $\pm 1$.
So the graphical representation will be as follows. We will denote the
critical points as dots. Then we will link the critical points $i$ and $j$
by a solid line if $|\mu_{ij}|=1$. We will not link them 
if $\mu_{ij}$ vanishes.
There will be at least one line coming from each critical point. (The
connectivity is quite analogous to Dynkin diagrams.)
So, if there are $k$ critical points, there will be a maximum of $k(k-1)/2$
BPS states and a minimum of $k-1$ BPS states since all critical points can
be connected through a sequence of BPS solutions \sv. 

2) {\it What determines the separatrix equation?}

Observe that
the topological charge associated with two critical points $i$ and $j$ is
\eqn\topcharge{T_{ij}=2e^{i\arg (W(z_j)-W(z_i))}
|W(z_j)-W(z_i)|}
and so is a complex number.
The mass $M$ of a domain wall is bounded by the absolute value of the
topological charge $T$:
\eqn\mt{M\geq |T|,}
and is saturated by a solution of the BPS equation. 
Now consider a situation where $i$ and $j$ and also $j$ and $k$ are
connected by a domain wall solution with BPS masses $M_{ij}$, $M_{jk}$ 
and topological
charges $T_{ij}$, $T_{jk}$. Let us consider the possibility of a BPS object
between $i$ and $k$. The possible BPS mass of such a solution is always bounded
by the following simple inequality:
\eqn\mttt{M_{ik} = |T_{ij}+T_{jk}|\leq |T_{ij}|+|T_{jk}| = M_{ij} +
M_{jk}.}
The inequality is saturated only when the phases of $T_{ij}$ and $T_{jk}$ are 
the same. When the equality \mttt\ is saturated, such that
$M_{ik}=M_{ij}+M_{jk}$, then the domain wall with mass $M_{ik}$ decays
into the two other domain walls. 
Since the phase of the topological charge
comes from the argument of the difference of the superpotential, we can
calculate the boundaries in the deformation parameter space where different
solitons are created or destroyed as we change the parameters. 
Each such boundary is determined by three critical points and determines
whether a solution between a particular pair of them becomes unstable or
not. The entire parameter space will therefore be divided into many
different domains and each domain will have the same number of possible BPS
solutions. 

3) To map the entire parameter
space we pick a point in the space where the BPS configuration is easily
determined. As we move across a boundary a certain state can be created (if
it was not there) or destroyed (if it was there).     
This
technique will be applied in the next section where we find the separatrix
curves for a general quintic superpotential. 

\newsec{Finding the BPS Configurations}

\subsec{Quartic Superpotential}
The simplest nontrivial superpotential 
is of course one with two critical points.
This allows a single BPS state and hence a single domain wall.
Next would be one which has three critical points. 
In this case of $k=3$ roots, and actually for all $k\geq 3$, one can argue
that {\it any} pair of roots can be connected through a sequence of domain
wall solutions \sv.
By rescaling and fixing
the value of the field we can fix two of the critical points to be, say at
$z_1=-1$ and $z_2=1$. The third critical point can be at an arbitrary point in
the complex plane, say at $z_3=\mu$ (this case is discussed in detail in \ab).
When $\mu$ is a real number, $\mu>1$,  
the critical points in the $W-$ plane will be
colinear and the only straight line connecting $z_1$ with $z_3$ will be
through $z_2$. So there can only be two types of domain walls. 
The same conclusion -- i.e. that there are only two BPS states -- 
can be drawn when $|\mu|<1$ for real $\mu$, and also for $\mu<-1$.  
Let us now see what happens when we move away from the real line, 
holding fixed $z_1$ and $z_2$,
for the case of $-1<\mu<1$.
As $\mu=\mu_1+i\mu_2$ ($\mu_1$, $\mu_2$ 
are real numbers) moves away from the real line, the number of BPS states stays
the same until we reach a boundary in the complex $\mu$ plane where a new
BPS state appears, arising from the domain wall between $z_1$ and $z_2$.
This boundary is defined by the condition that the phases of 
the topological charges
$T_{13}$ and $T_{32}$ are the same \ab.
When the phases are the same then the inequality of the masses saturate and
we have $M_{12}=M_{13}+M_{32}$ ($M_{ij}$ is the mass of the soliton
connecting the roots $z_i$ and $z_j$). 
Similar boundaries can be found for the initial cases of $|Re(\mu)|>1$,
determined by the equality of masses of $M_{12}$ and $M_{23}$ or $M_{31}$
and $M_{12}$. 

There is a reflection symmetry of the boundaries in the real
line. These three boundaries together form the  separatrix curve 
and the equation
can be written down as the following condition on the real and imaginary
parts of $\mu$:
\eqn\sepaforab{3\mu^4_1 + 2\mu^2_1 \mu^2_2 -\mu^4_2 -
6\mu^2_1 -6\mu^2_2+3=0.}
Note that this equation {\it does not distinguish} which
BPS state melts away as we cross a boundary. Different branches 
of eq. \sepaforab\
will correspond to one of the boundaries which we discussed above, obtained
from the relations between the possible masses of domain walls.
So if we put $F_{ijk}\equiv M_{ij}+M_{jk}-M_{ik}$, the separatrix equation 
will be equivalent to $F_{123}F_{132}F_{312}=0$, after some algebraic 
manipulations.
This observation will be quite crucial in identifying various BPS states in
the cases with more than three critical points.
The real line will appear as a solution of the separatrix equation, but it
will be a line of marginal stability, so the number of BPS states do not
change as we cross the real line. The connectivity of the roots for the
quartic superpotential is therefore very simple: either any root is
connected to any other root (for a total of three BPS states), 
or two of the roots are not directly connected 
(for a total of two BPS states). 
This result is given in Fig. 4 of Ref.\ab.
The connectivity signals a {\it possible}  BPS state. It can be occupied or
be vacant. 
Now when the occupied BPS states are  such that we have an
enclosed domain, then we have a junction of the three domain walls and
a 1/4 BPS state. If they do not enclose a separate domain, say just 
two of the edges of a triangle, then we have two BPS domain walls which
never join and the whole configuration will be 1/2 BPS.
So when the positions of the critical points are more or less colinear (in W-space)
domain wall junctions do not develop.
This can be used in the cases with more than three critical points, where 
the positions of three particular ones will more or less follow the
pattern described above, although the very existence of the other critical
points do interfere with the detailed shape of the separatrix curves.

\subsec{Quintic Superpotential}

Next we analyze the $D=4$ W-Z model with a general quintic
superpotential, 
\eqn\quintic{W=z^5+\sum_{i=1}^4 \alpha_i z^i,}
where $\alpha_i$, $i=1,\ldots, 4$ are complex deformation parameters.
Critical points of the superpotential are points $z_a$ where $dW(z_a)=0$.
The possible connections of the critical points in this case are shown in
Figure 1, where each dot represents a root $i$ (or critical point) 
and each line represents a
possible domain wall solution interpolating between critical points $i$ 
and $j$. When such a line exists between two given roots, 
$|\mu_{ij}|=1$, and it thus represents a possible BPS state. When there is no 
line between two given roots $\mu_{ij}=0$ and there can be no BPS state. 
Therefore it is easy to see that Figure 1 exhausts all possible connections
of critical points. 
So one expects that in some domain the number of BPS states is the smallest
possible, namely three (as in Figs 1--A,-- B), while in some other domain the
maximum number of BPS states, namely six (as in Figure 1--F) is obtained,
depending on the choice of superpotential, i.e. deformation parameters 
$\alpha_i$. 
However, Figure 1--C deserves some further comments. In the following we will
see that in no finite domain of deformation parameters will the
connectivity be as in C. 
This is actually easy to understand geometrically. In such a four-gon -- defined
by roots $i,j,k$ and $l$ -- are $i'j'k'$ connected for any cyclic permutation of
the four roots but $i'$ and $k'$ are not connected. So all the angles of
$i'j'k'$ has to be at least 90$^\circ$. But the sum of the angles of the
4-gon is 360$^\circ$ and we have a contradiction. 
What about the case of $k>4$ number of critical points?    
For a $k$-gon, the sum of angles is $(k-2)\cdot 180^\circ$. For a
configuration with no ``internal'' BPS solitons the sum of angles should be
at least $k\cdot 90^\circ$.  For $k\geq 5$ one might have such domains.
\vskip 0.2truein
\centerline{\epsfxsize 4.5truein \epsfysize 2.5truein\epsfbox{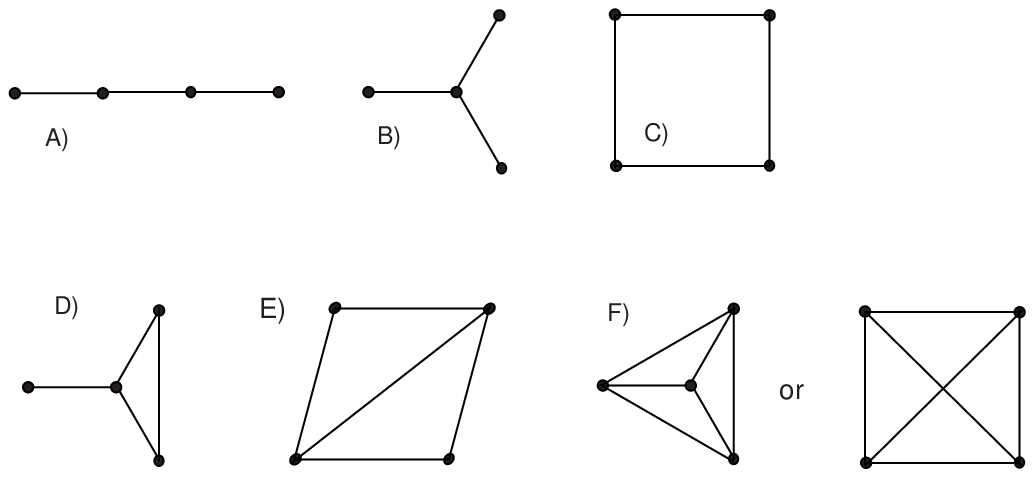}}
\noindent{\ninepoint\sl \baselineskip=8pt {\bf Fig.1}: {\rm
Possible connectivities of four critical points in the case of a 
quintic superpotential.}}
\bigskip

\noindent The soliton structure of 
any such massive deformation of a conformal theory
is characterized by the matrix $S=1-A$, where $A$ is an upper triangular matrix
whose elements $A_{ij}$ for $i<j$ are exactly $A_{ij}=\mu_{ij}
=\Delta_i\circ\Delta_j$ \cv.
However, this matrix does not in itself classify the possible junctions.
Precisely for this reason will Figure 1--F need some further comments. {\it
If} the actual location of the critical points is as in Figure 1--F
(triangle), 
then one can obtain a junction of domain walls by occupying all 
six states (this
junctions will look like a circle with three legs coming out). But imagine
that the locations of the critical points are as in Figure 1--F 
(square) with the
inclusion of the two BPS states connecting diagonal corners. Occupying all
these states would not give rise to a stable junction.   
  
In Eq. \quintic\ we fix two of the critical points to be at $z=\pm 1$, 
so that the four critical points are located at 
\eqn\critp{z_1=-1,\ \ z_2=1, \ \ z_3=\mu, \ \ z_4=\lambda,}
corresponding to the superpotential which takes the following form:
\eqn\wquint{W=z^5-{5\over 4}(\mu+\lambda)z^4+{5\over 3}(\mu\lambda-1)z^3
+{5\over 2}(\mu+\lambda)z^2-5\mu\lambda z.}
We thus have two complex parameters $\mu$ and $\lambda$ to vary, and in
general it is not easy to visualize different domains in this space of parameters.
A systematic way is to fix one of the complex parameters, say $\mu$ 
and have a sliced view of the separatrix walls.
We will consider a few representative values of $\mu$: 1) the case where three
points $z=-1, z=1,z=\mu$ are at vertices of an equilateral triangle,
(This includes the case we already discussed in the introduction
which corresponds to the situation where the fourth critical point is at
 the center of the triangle. For this case we already know the possible
connectivities of the critical points and we can use it as the `initial
data' for our analysis.) 
 2) the case where three
points are colinear on the real axis and finally 3) the case which 
includes the ${\bf Z}_4$ symmetric case.

For a generic configuration of roots (i.e. when $z_3$ is not colinear with
$z_1$ and $z_2$) one can obtain the complete set of separatrix curves as
follows. Pick any two roots $z_i$, $z_j$ ($i>j$) and consider the basic
separatrix curve joining them as defined by the equation
$F_{4ij}F_{4ji}F_{j4i}=0$. Then the condition that the product of all these
groups of terms vanishes is the equation for the ``complete'' separatrix
curve, just as it is in the case of a single pair of roots when we have a
quartic superpotential.
Now we will focus on the three 
cases. In the first case we take the three fixed roots to
be at the vertices of an equilateral triangle, i.e. 
$z_1=-1,z_2=+1,z_3=i\sqrt{3}$ (see Figure 2).  

\centerline{\epsfxsize 4truein \epsfysize 4.2truein\epsfbox{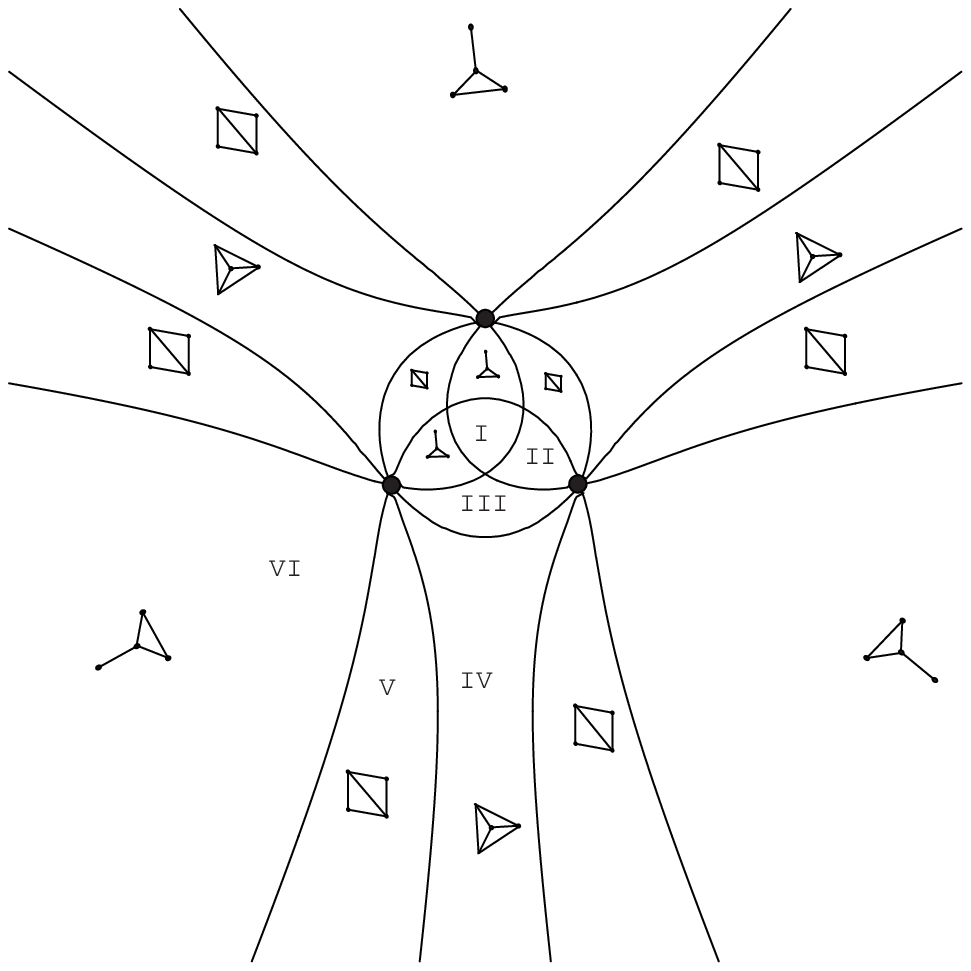}}
\noindent{\ninepoint\sl \baselineskip=8pt {\bf Fig.2}: {\rm
${\bf Z}_3$--symmetric case. Connectivity of roots depending on the value of 
$z_4=\lambda$. The three fixed roots are located at $z_1=-1$, $z_2=1$ and 
$z_3=i\sqrt{3}$.}}
\bigskip

\noindent In this case there is a ${\bf
Z}_3$--symmetry generated by rotations of $2\pi/3$ in the center of the
triangle. In the second case we take
the roots to be colinear $z_1=-1,z_2=+1,z_3=+3$ (see Figure 3). 
In this case there is a  ${\bf Z}_2$ symmetry
generated by reflections along the vertical line $\lambda_2=0$.  
The last configuration is where $z_1=-1, z_2=+1$ and $z_3=-1+2i$ (see
Figure 4) and so contains the ${\bf Z}_4$-symmetric superpotential 
(for $z_4=1+2i$) which has been much studied.

Before going into details with the different phase diagrams and 
determining in which domains we have how many BPS states and so forth, 
we start with a global view (i.e. far away from the origin).
What determines the angles between the curves of marginal stability?
For that we will take a long-distance view of the separatrix curves. This
limit corresponds to both $\lambda_1$ and $\lambda_2$ large. 
For any fixed value $z_3=\mu$, one can write down the separatrix equation 
as a sixth order equation in $\lambda_1$ and $\lambda_2$
(for example for the pair of roots $(z_2,z_4)$ and $(z_4,z_3)$ as follows):
\eqn\septrix{\eqalign{0= & -{5\over 12}\lambda_1^5\lambda_2(-1+\mu)^3(3+\mu)+{5
\over 6}\lambda_1^4\lambda_2(-1+\mu)^3(1+3\mu+\mu^2)\cr 
& -{5\over 12}\lambda_1^3\lambda_2(-1+\mu)^3(-5+\mu+3\mu^2+\mu^3)\cr 
& -{5\over
18}\lambda_1^2\lambda_2(-1+\mu)^3(6+18\mu+6\mu^2+\lambda_2^2(6+3\mu+\mu^2))\cr
& -{1\over
18}\lambda_2(-1+\mu)^3(15\mu^2(2+\mu)+\lambda_2^4)(-3+6\mu+2\mu^2)+5\lambda_2^2(-2+9\mu+3\mu^2))\cr
& +{5\over 36}\lambda_1\lambda_2(-1+\mu)^3(3\lambda_2^4(3+\mu)+3\mu(8+9\mu+3\mu^2)
+\lambda_2^2(15+13\mu+9\mu^2+3\mu^3)).
}}
The sliced view of this separatrix equation will be shown in Figure 2--4 for 
particular values of $\mu$ mentioned above.
The angle between the lines of marginal stability (corresponding to two
roots $z_a, z_b$) and the line $\lambda_1=0$ is clearly determined by
the fraction $\rho=\lambda_1/\lambda_2$. 
So by dividing the above equation with $\lambda_2^6$ and taking the limit
$\lambda_1,\lambda_2$ large we obtain:
\eqn\angles{0=-{5\over 12}\rho^5(-1+\mu)^3(3+\mu)+
{5\over 36}3\rho(-1+\mu)^3(3+\mu),}
which has the real solutions $\rho=\pm 1$. So far away, the lines meet at
an angle of $\pi/2$. The same is the case in the $k=3$ theory, where the curves of
marginal stability (for the ``basic'' separatrix curve discussed in section
5.1) meet at an angle $\pi/2$ at infinity. Now consider the 
${\bf Z}_3$ symmetric case as in Figure 2.
For any pair of roots $(z_a,z_b)$ we have a basic separatrix curve joining
them. Far away from the origin these curves meet at an angle
of $\pi/2$. Now, because of the ${\bf Z}_3$--symmetry, the angle between two
neighboring curves must then be $(\pi/2)/3=\pi/6$. Asymptotically we
therefore have 12 domains. 

We start by counting the number of possible BPS states for 
the ${\bf Z}_3$-symmetric configuration of roots, see Figure 2. 
Generally we will
call $z_1$ as root $1$, $z_2$ as root $2$ and so on.
We start with the
most symmetrical configuration, where the fourth root $\lambda$ is in the
center of the triangle defined by the roots 1, 2 and 3. We call this small
domain {\it I}. {\it I} is defined as the intersection of three domains:
one where 1 is connected to 4 and 4 is connected to 3, but 1 and 3 is {\it
not} connected (this comes from the basic separatrix curve connecting 1 and
3), one where 2 is connected to 4 which is connected to 3, but 2 and 3 is
not connected and finally one where 1 is connected to 4 and 4 is connected
to 2 but 1 and 2 are not connected. This shows that the connectivity of the
diagram in domain {\it I} must be of type B. The number of possible BPS
states is therefore 3. The number of BPS states in the other domains can
now be determined by using the rules described in the last section in
crossing the different separatrix curves.

{\bf I}$\rightarrow${\bf II}: 
In going to domain {\it II} one crosses the
line $F_{143}=0$ and since there was no connecting between 1 and 3 to start with
these two roots gets connected by a BPS solution. The number of BPS states
in {\it II} is then 4 and the connectivity is of type D. 
{\bf II}$\rightarrow${\bf III}:
In going to domain {\it III} one crosses the line $F_{243}=0$ and since 
there was no connecting between 2 and 3 to start with
these two roots gets connected by a BPS solution. The number of BPS states
in {\it III} is then 5 and the connectivity is of type E.
{\bf III}$\rightarrow${\bf IV}: 
In going to domain {\it IV} one crosses the line $F_{142}=0$ and since 
there was no connecting between 1 and 2 to start with 1
and 2 to will be connected such that all roots are connected and the number
of BPS states is 6. The connectivity is of type F.
{\bf IV}$\rightarrow${\bf V}: 
In going to domain {\it V} one crosses the line $F_{314}=0$ and since 
there was a connecting between 3 and 4 to start with, this domain wall
disappears and instead 1 and 2 is connected. The connectivity is then of
type E with 5 possible domain walls. 
{\bf V}$\rightarrow${\bf VI}: 
In going to domain {\it VI} one crosses the line $F_{124}=0$ and since 
there was a connecting between 1 and 4 to start with, this domain wall
disappears. The connectivity is then of type D with 4 possible BPS states.
By ${\bf Z}_3$-symmetry this analysis determines the possible number of BPS
states in all domains and hence we have a complete determination of the
possible domain walls and junctions for a potential with this particular 
symmetry. For this class of superpotentials, the number of BPS states
varies from three to six.

A similar analysis can be carried out for the ${\bf Z}_2$--symmetric case
in Figure 3. 
When we simply plot the corresponding 
Eq. \septrix\ for all pairs of roots
then we get more lines than is shown in
Figure 3. However, some of these lines are lines of marginal stability, just
like the real axis is for a quartic superpotential as discussed in section
5.1, and should be ignored.
 
\centerline{\epsfxsize 4truein \epsfysize 4.2truein\epsfbox{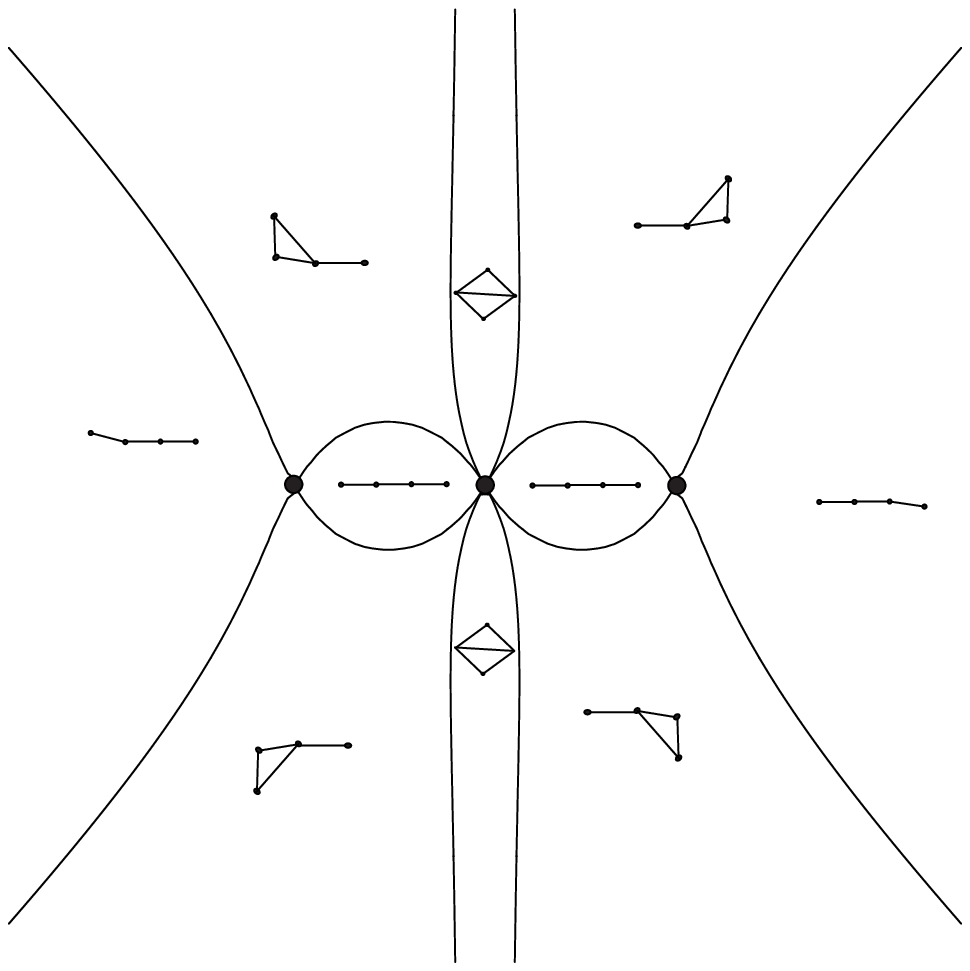}}
~\noindent{\ninepoint\sl \baselineskip=8pt {\bf Fig.3}: {\rm
${\bf Z}_2$--symmetric case. Connectivity of roots depending on the value of 
$z_4=\lambda$. The three fixed roots are 
located at $z_1=-1$,$z_2=1$ and $z_3=3$.}}
\bigskip

\noindent However, here the three fixed roots are all colinear so the
resulting diagram is very simple. To determine the possible BPS states in
the different domains, one can start by taking $1<z_4<3$ and real. In this
case the configuration is known \sv: all roots are successively connected as
shown in Figure 3. The configuration in other domains is then 
simply determined by
crossing the different curves of marginal stability.  
For this case the number of BPS states varies from three to five.

The case including the ${\bf Z}_4$--symmetric potential is presented in
Figure 4. At first glance this figure looks very complicated. However, it
has some features common with Figure 2. For example, eight separatrix lines
emanate from each critical point. For this choice of parameters, the number
of BPS states varies from three (around the 'center') to six (at the ${\bf
Z}_4$ symmetric point for example). 
So all possible connectivities are realized,
except the minimal case of three BPS states (Figure 1--A) 
and Figure 1--C of course.  

We have found all the possible BPS states. Now let us focus on
the junction configurations. As mentioned before, a triangle leads to a
junction of three domain walls.
If only one edge of the triangle is occupied, it is a 1/2 BPS
state of a single domain wall. 
When all three edges are occupied, then all the tensions will be balanced
and this will lead to a 1/4 BPS configuration of junctions of domain walls.
More generally, one could have junctions of any number of domain walls.
Here is how we can define a junction configuration in this case.
First find the locations of the critical points in the $W$-space.
The integral curves will be straight lines between two critical points
$i,j$ in this space, and will have
corresponding `soliton' number $\mu_{ij}$, which can also be zero.
Next pick any number of critical points in the $W$-space, 
such that the successive connection of these points
form a convex polygon. If all the edges have nonvanishing $\mu_{ij}$, that is, if
the polygon is closed then we have a nontrivial junction, and the 
inside of the
polygon will have 1/4 supersymmetry. Each of the edges of the polygon will
have 1/2 of the original supersymmetry, and only at the vertices, that is
at the critical point, is all of the original supersymmetry preserved.
Again, we see in this `graphical' understanding of various SUSY
configurations that there is no room for 3/4 BPS states 
in the W-Z model \gght.

\centerline{\epsfxsize 4truein \epsfysize 4.2truein\epsfbox{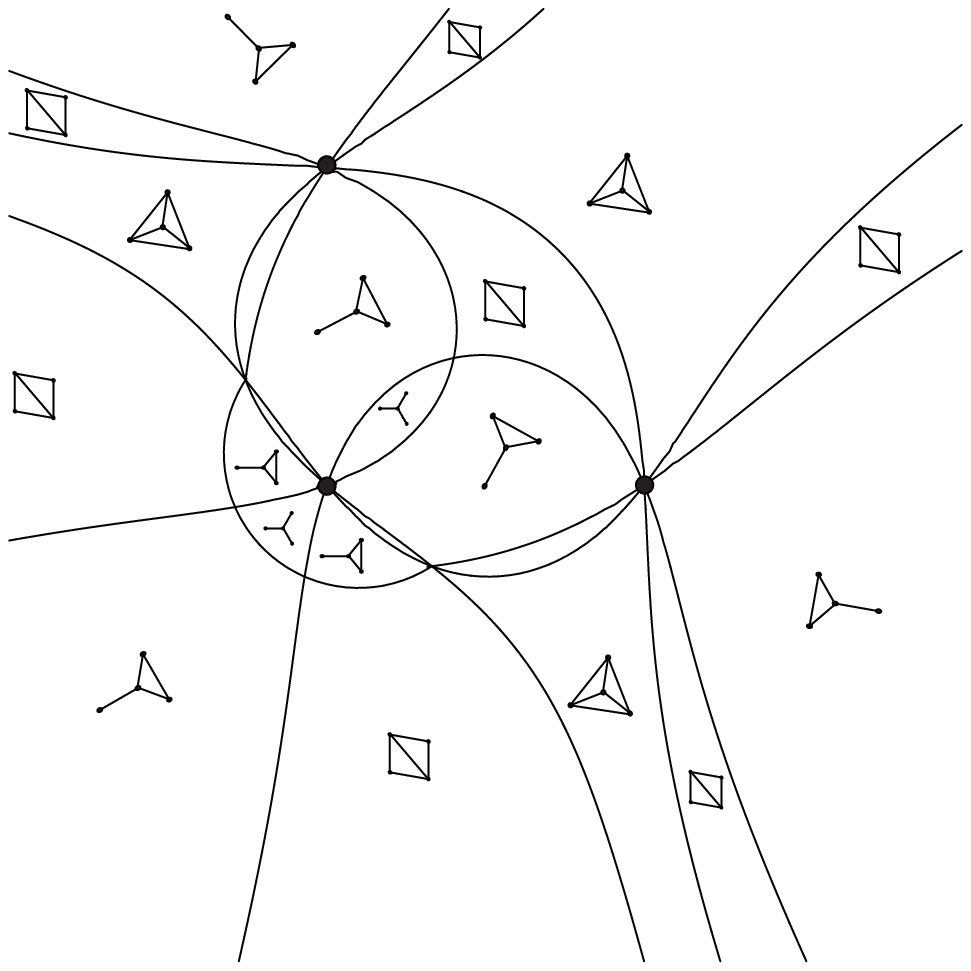}}
\noindent{\ninepoint\sl \baselineskip=8pt {\bf Fig.4}: {\rm
Connectivity of roots depending on the value of 
$z_4=\lambda$. The three fixed roots are 
located at $z_1=-1$, $z_2=1$ and $z_3=-1+2i$. In the empty domains the
connectivity is of type E.}}
\bigskip

\noindent If any of the $\mu_{ij}$ along the edge of the polygon is zero, then we
cannot define the `inside' of the polygon and there will be no junction.
We will have just domain walls with extend to infinity (in the coordinate
space) and which never join. 
The number of preserved supercharges will be two.

\newsec{Discussion}

So far we have been discussing the possible BPS states and junctions in 
the W-Z model. We have summarized our result in Figures 2--4 where we can
easily read off the number of BPS states as well as possible 
BPS junction configurations for a given deformation parameter.
So what is the use of all this? First of all, we have used a method general
enough to be utilized for counting BPS states for other types of
superpotentials. 
Secondly, the BPS data of W-Z models (or those with other
superpotentials) which can be obtained from higher dimensional theories
will reflect the BPS data of the original theories. 

Apart from these practical things, we would also like to point out some of the 
possible connections to works done in the context of string 
compactifications and also brane configurations.
\lref\toric{N.C. Leung and C. Vafa, {\it Branes and Toric Geometry},
Adv. Theor. Math. Phys. {\bf 2} (1998) 91, hep-th/9711013.}
Due to the relation to Calabi-Yau compactifications we can reinterpret
our results as that of counting numbers of BPS $D$-branes wrapped 
around supersymmetric
cycles. On top of each domain wall there is a `sphere' wrapping around a
supersymmetric cycle, whose radius vanishes at the critical points.
This is reminiscent of toric geometry:
We have vanishing spheres at the
critical points and have finite radius cycles over the line interpolating
two critical points. That is spheres in the internal dimension over the
domain walls will be revealing some of the structures of Calabi-Yau spaces. 
In particular, it has been shown that certain toric geometries, 
which has vanishing cycles, can be translated into a brane configuration \toric.
Thus another very interesting application comes from the $T$-dual picture of the
Calabi-Yau compactifications, i.e. the brane configurations.
As an example consider the following situation.
The brane configuration for ${\cal N}=1$ $SU(N_c)$ supersymmetric YM is given 
in Type IIA string theory
with $N_c$ $D4$ branes extending between two sets of coincident $NS5$
branes as follows. With the coordinates  
\eqn\coordinates{s=x^6 +i x^{10},\  v=x^4+ix^5,\ w=x^8+ix^9.}
the $D4$ brane is located at $v=w=0$ and extended in the $s$-direction.
The $NS5$ brane is located at $s=w=0$ and is extended in the 
$v$-direction, and the $NS5'$-brane is at
$v=0,s=L$ and is extended along the $w$-direction. 
Consider a configuration of $m$ coincident $NS5$ branes connected by $N_c$
$D4$ branes to $m'$ coincident 
$NS5'$ branes. There will be two adjoint superfields $\Phi, \Phi'$, which
describe fluctuations of the fourbranes in the $w$ and $v$ directions
respectively, whose classical superpotential is 
\eqn\coincidentNS{W={a\over m+1}Tr\Phi^{m+1} +{a'\over m'+1}Tr\Phi^{m'+1}+\cdots.}
Imagine having the $k$ $NS5$ branes in the $(x^8,x^9)$ 
plane at $k$ different points $w_j$.
Since the $\{ w_j\}$ correspond to locations of heavy objects they appear
as parameters rather than moduli in the gauge theory description and give
rise to a polynomial superpotential for $\Phi$ where 
$W'(\Phi) = a\prod^{m}_{j=1} (\Phi-w_i)$. This shows how superpotentials of
the form discussed in this paper can arise from brane configurations.

Another very interesting result can be obtained with 
similar methods in the study of BPS states of 
Argyres and Douglas superconformal theories \argyresdouglas\apsw\ehiy\ , 
as in Ref.\sv. 
In fact, if we consider a degenerate choice of polynomial, where
$P_m=(dW/dx)^2$, the problem becomes identical to the problems we have
discussed here. Exact equations for the separatrix curves 
can be obtained but will be quite
complicated and involve certain elliptic functions. 

As discussed in section 3, 
when we consider Type IIA string theory compactified on a Calabi--Yau fourfold 
we obtain a 1+1
dimensional effective theory which gives the vacuum structure and the D4
branes wrapping around the supersymmetric cycles give solitons
interpolating the vacua. 
If we start with $M$-theory, which is the strong coupling regime of the
Type IIA theory, we end up with an effective 2+1 dimensional theory, 
with similar vacuum and
domain wall structure. However, there is something more. Due to one more space 
dimension, the vacua can arrange such that there can be junctions of the
domain walls. From the point of view of string theory this
extra dimension is a nonperturbative effect. 
Thus having a full understanding of lower--dimensional integrable models  
might not guarantee an understanding of higher--dimensional integrable
model, just as understanding fully perturbative field theory does not
guarantee any insight into a fully nonperturbative field theory. 

The superpotential we have studied in this paper is the simplest kind
involving only one type of field. There are many extensions that can be
made with multiple species of fields. 
One nice extension would be the study of the 
$D-E$ series \ehiy\ of singularities
and the corresponding  W-Z models.
In the case of W-Z models of $A_n$ type, one always has a 
single type of domain walls
between two critical points, because there is only one type of complex scalar
field in the theory. However, if we have multiple species of scalar fields we 
have the possibility of multiple types of domain walls between the critical
points. It would be interesting to generalize the method used here to study these
systems and also find junctions of multiple species of branes.
Theories such as the $CP^n$ models have multiple species of domain walls between
critical points, which can be labeled by a group theory index. 
So when we consider junctions of a multiple of these domain walls, perhaps
only a certain combinations will lead to a BPS junction. 
This certainly deserves a further study.

There are still some open questions, we would like to answer in the near future:
How do we describe junctions of domain walls in the higher--dimensional
Calabi--Yau geometry? Are stable junctions classified by some topological
class, related to the higher--dimensional geometry?

\bigskip
{\bf Acknowledgements}

We have benefitted from useful conversations/correspondence with B. Andreas, 
K. Hori, B. Pioline, E. Witten and 
especially critical and helpful comments of C. Vafa. The work of
S.N. was supported by BK21 (Brain Korea 21) Program of Korea Reseach
Foundation and that of K.O. by the Danish Natural Science Research Council.

\vfill\eject
\listrefs
\end